\documentclass[aps,prb,reprint,superscriptaddress]{revtex4-1}
\usepackage{geometry}
\usepackage{setspace}
\usepackage{fullpage}
\usepackage{color}
\usepackage{hyperref}
\usepackage{graphicx}
\begin{document}

\title{FABRICATION OF GRAIN BOUNDARY JUNCTIONS USING NdFeAs(O,F) SUPERCONDUCTING THIN FILMS}

\author{Taito\ Omura}\email{oomura.taito@j.mbox.nagoya-u.ac.jp}
\affiliation{Department of Crystalline Materials Science, Nagoya University, Furo-cho, Chikusa-ku, Nagoya 464-8603, Japan}\author{Takuya\ Matsumoto}
\affiliation{Department of Materials Physics, Nagoya University, Furo-cho, Chikusa-ku, Nagoya 464-8603, Japan}
\author{Takafumi\ Hatano}
\affiliation{Department of Crystalline Materials Science, Nagoya University, Furo-cho, Chikusa-ku, Nagoya 464-8603, Japan}
\affiliation{Department of Materials Physics, Nagoya University, Furo-cho, Chikusa-ku, Nagoya 464-8603, Japan}
\author{Kazumasa\ Iida}
\affiliation{Department of Crystalline Materials Science, Nagoya University, Furo-cho, Chikusa-ku, Nagoya 464-8603, Japan}
\affiliation{Department of Materials Physics, Nagoya University, Furo-cho, Chikusa-ku, Nagoya 464-8603, Japan}
\author{Hiroshi\ Ikuta}
\affiliation{Department of Crystalline Materials Science, Nagoya University, Furo-cho, Chikusa-ku, Nagoya 464-8603, Japan}
\affiliation{Department of Materials Physics, Nagoya University, Furo-cho, Chikusa-ku, Nagoya 464-8603, Japan}
\date{\today}

\begin{abstract}
We report on the growth of NdFeAs(O,F) thin films on [001]--tilt MgO bicrystal substrates with misorientation angle $\theta_{\rm GB}=6^\circ, 12^\circ, 24^\circ$ and $45^\circ$, and their inter- and intra-grain transport properties. X-ray diffraction study confirmed that all our NdFeAs(O,F) films are epitaxially grown on the MgO bicrystals. The $\theta_{\rm GB}$ dependence of the inter-grain critical current density $J_{\rm c}$ shows that, unlike Co-doped BaFe$_2$As$_2$ and Fe(Se,Te), its decay with $\theta_{\rm GB}$ is rather significant. As a possible reason of this result, fluorine may have diffused preferentially to the grain boundary region and eroded the crystal structure.
\end{abstract}

\maketitle

\section{Introduction}
It is very important to examine the current-limiting effect of grain boundaries (GBs) for fabricating superconducting wires and tapes based on high critical temperature ($T_{\rm c}$) superconductors. For cuprate superconductor YBa$_2$Cu$_3$O$_{7-\delta}$ (YBCO), the $J_{\rm c}$ characteristic across the grain boundary ($J_{\rm c}^{\rm inter}$) has been intensively investigated using a bicrystal substrate in which two single crystals with different orientations are thermally jointed.\cite{Hilgenkamp} Although $J_{\rm c}^{\rm inter}$ of YBCO does not decrease with increasing the misorientation angle $\theta_{\rm GB}$ until a critical angle $\theta_{\rm c}$ ($3^\circ$$\sim$5$^\circ$), $J_{\rm c}^{\rm inter}$ attenuates exponentially with $\theta_{\rm GB}$ above $\theta_{\rm c}$. Therefore, for practical use as superconducting tapes, biaxially textured YBCO conductors whose in-plane orientation distribution is smaller than $\theta_{\rm c}$ are required. However, this demands a costly fabrication process and delayed significantly conductor applications.

Iron-based superconductors (FBS) form a material class that has a high $T_{\rm c}$ next to the cuprates and a high upper critical field ($H_{\rm c2}$) exceeding 100\,T with a moderate $H_{\rm c2}$ anisotropy.\cite{Putti} Regarding the grain boundary characteristics, Katase $et$ $al$. reported that $\theta_{\rm c}$ is as large as $9^\circ$ for Co-doped BaFe$_2$As$_2$, which is about twice larger than that of YBCO.\cite{Katase} In addition, the attenuation of $J_{\rm c}^{\rm inter}$ above $\theta_{\rm c}$ was more moderate than YBCO. As a result, $J_{\rm c}^{\rm inter}$ of Co-doped BaFe$_2$As$_2$ was higher than that of YBCO at 4\,K and $\theta_{\rm GB}\geq 20^\circ$. Our group reported that P-doped BaFe$_2$As$_2$ showed also a slow decay of $J_{\rm c}^{\rm inter}$ with $\theta_{\rm GB}$ and a high $J_{\rm c}^{\rm inter}$ of 10$^6$\,A/cm$^2$ for $\theta_{\rm GB}=24^\circ$ at 4\,K, which is the highest value ever reported for YBCO and FBS for this misorientation angle.\cite{Sakagami} Similar to the BaFe$_2$As$_2$ system, the transition from strong-link to weak-link occurs at $\theta_{\rm c}$$\sim$9$^\circ$ for Fe(Se,Te) grown on SrTiO$_3$ bicrystals.\cite{Si} All these results are very promising for high field conductor applications.

\begin{figure}
	\centering
		\includegraphics[width=\columnwidth]{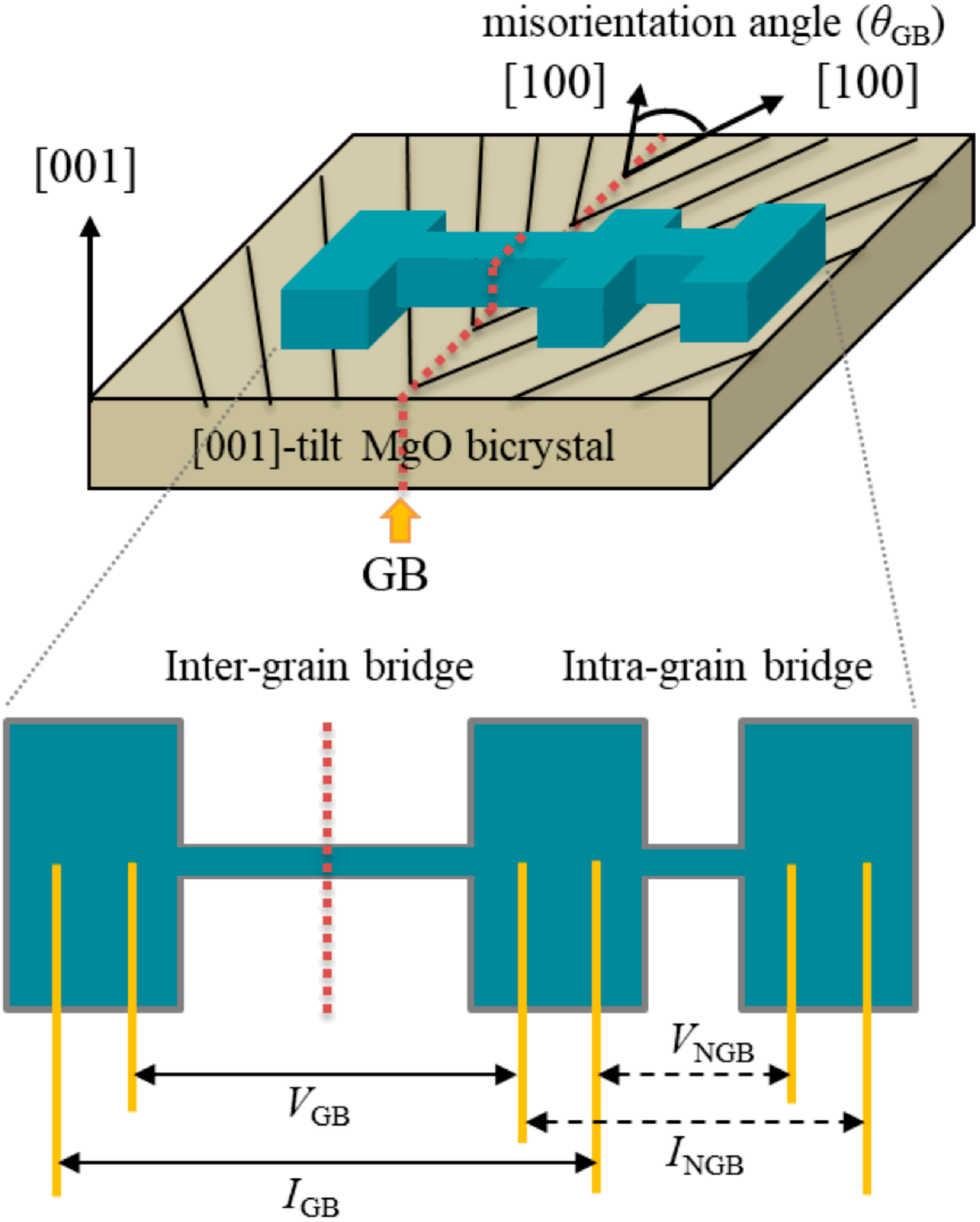}
		\caption{A schematic illustration of the micro-bridge for transport measurements. $V_{\rm GB}$ and $I_{\rm GB}$ are the voltage and current contacts for the GB junction, while $V_{\rm NGB}$ and $I_{\rm NGB}$ are those for the intra-grain bridge.}
\label{fig:figure1}
\end{figure}

Among the various FBSs discovered to date, $Ln$Fe$Pn$(O,F) ($Ln$:\,lanthanoid, $Pn$:\,pnictogen, and 1111) has the highest $T_{\rm c}$ up to 58\,K,\cite{Miyazawa, Fujioka} indicative of potentially high $J_{\rm c}$. However the characteristics of grain boundary has not yet been examined in detail since the crystal growth of this system is very difficult. We have succeeded in growing high quality epitaxial NdFeAs(O,F) thin films by molecular beam epitaxy (MBE).\cite{Kawaguchi-1, Kawaguchi-2} This opened the opportunity to fabricate 1111 films on bicrystal substrates. Here, we report on the transport properties of NdFeAs(O,F) films grown on bicrystal substrates with various misorientation angles.

\section{Experimental details}     
NdFeAs(O,F) thin films were grown by MBE. Solid sources of Fe, Fe$_2$O$_3$, As, NdF$_3$, and Ga were charged in Knudsen cells. Here, Ga was used as a F-getter to control the amount of fluorine.\cite{Kawaguchi-3} The films were grown on 6$^\circ$, 12$^\circ$, 24$^\circ$ and 45$^\circ$ [001]--tilt MgO bicrystal substrates and the substrate temperature was fixed at 800$^\circ$C. We have first grown a NdFeAsO parent phase layer, and then NdOF subsequently. This resulted in a superconducting 1111 film with a high $T_{\rm c}$ as fluorine diffused into the 1111 phase from the top fluoride layer\,\cite{Kawaguchi-3, Uemura}. The film thickness of NdFeAsO was about 160\,nm and that of NdOF about 50\,nm. Reflection high energy electron diffraction (RHEED) was used for $in$--$situ$ monitoring the surface structure during the thin film growth. Phase purity and crystalline quality of the obtained thin films were examined by X-ray diffraction (XRD) using Cu--K$_\alpha$ radiation. Microstructural characterization was performed by transmission electron microscope (TEM).

After the NdOF over-layer was removed by Ar-ion milling, the films were photolithographically patterned and etched by Ar-ion milling to form micro-bridges for transport measurements. As shown in Figure\,\ref{fig:figure1}, the films were patterned so as to measure the transport properties of both grain boundary and intra-grain on the same sample. The bridges were 20--40\,$\mu$m-wide and 1\,mm--long across the grain boundary and 0.25\,mm--long within the grain. The current--voltage characteristics and the temperature dependence of resistivity were measured by a four-probe method. A criterion of 1\,$\mu$V/cm was employed for determining intra-grain $J_{\rm c}$. On the other hand, inter-grain $J_{\rm c}$ was defined as the intersection between $E=0$ and a linear fit to the non-ohmic linear differential region (i.e., the region where the electric field $E$ depended linearly on the current density $J$).
     
\begin{figure*}[ht]
	\centering
		\includegraphics[width=\columnwidth]{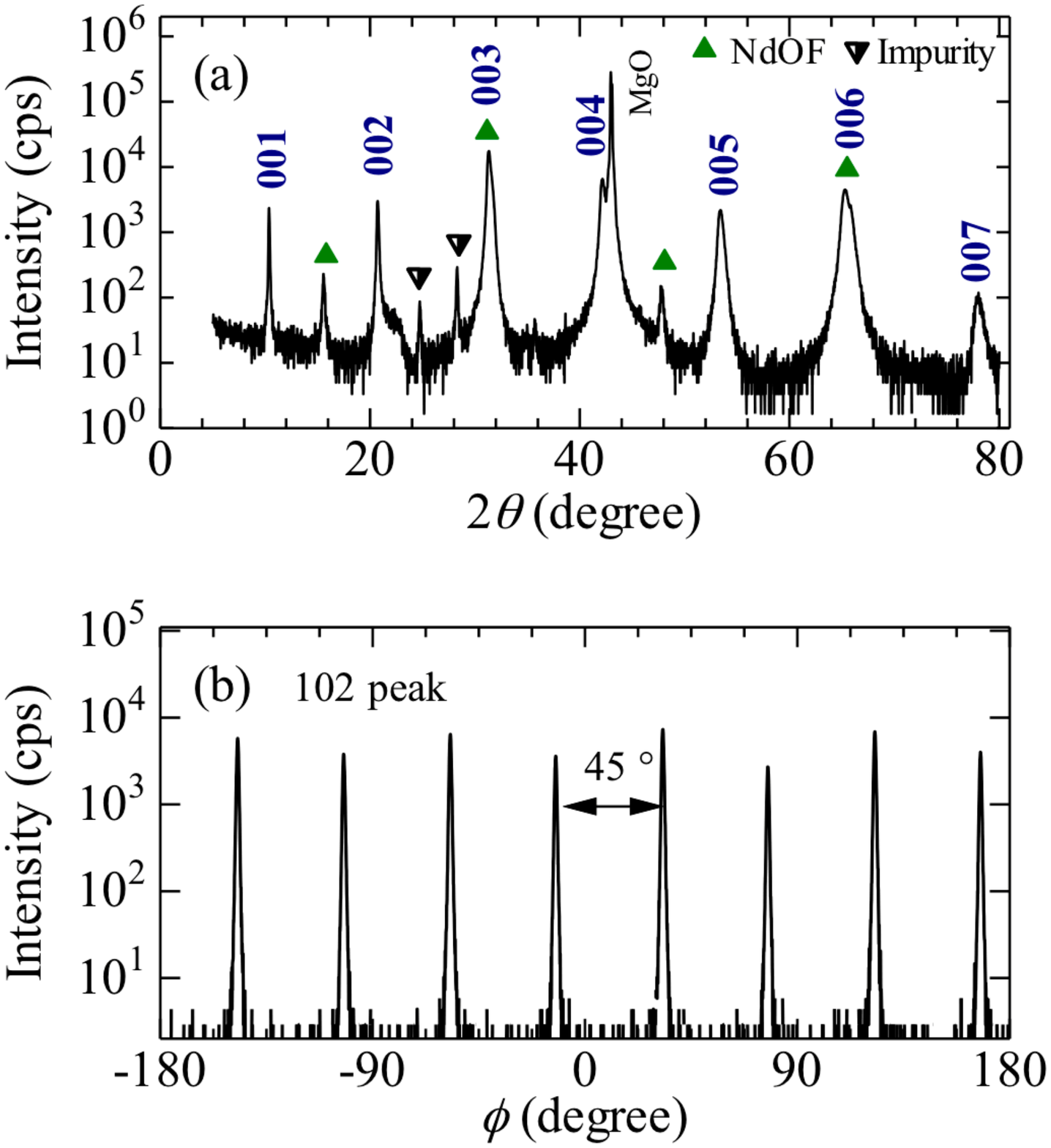}
		\caption{(a) $\theta$--2$\theta$ scan of the NdFeAs(O,F) film on [001]--tilt MgO bicrystal with $\theta_{\rm GB}=45^\circ$ measured with the Bragg-Brentano geometry using Cu--K$_\alpha$ radiation. (b) $\phi$--scan of the 102 NdFeAs(O,F) peak.}
\label{fig:figure2}
\end{figure*}

\section{Results and discussion}

\begin{figure*}[ht]
	\centering
		\includegraphics[width=12cm]{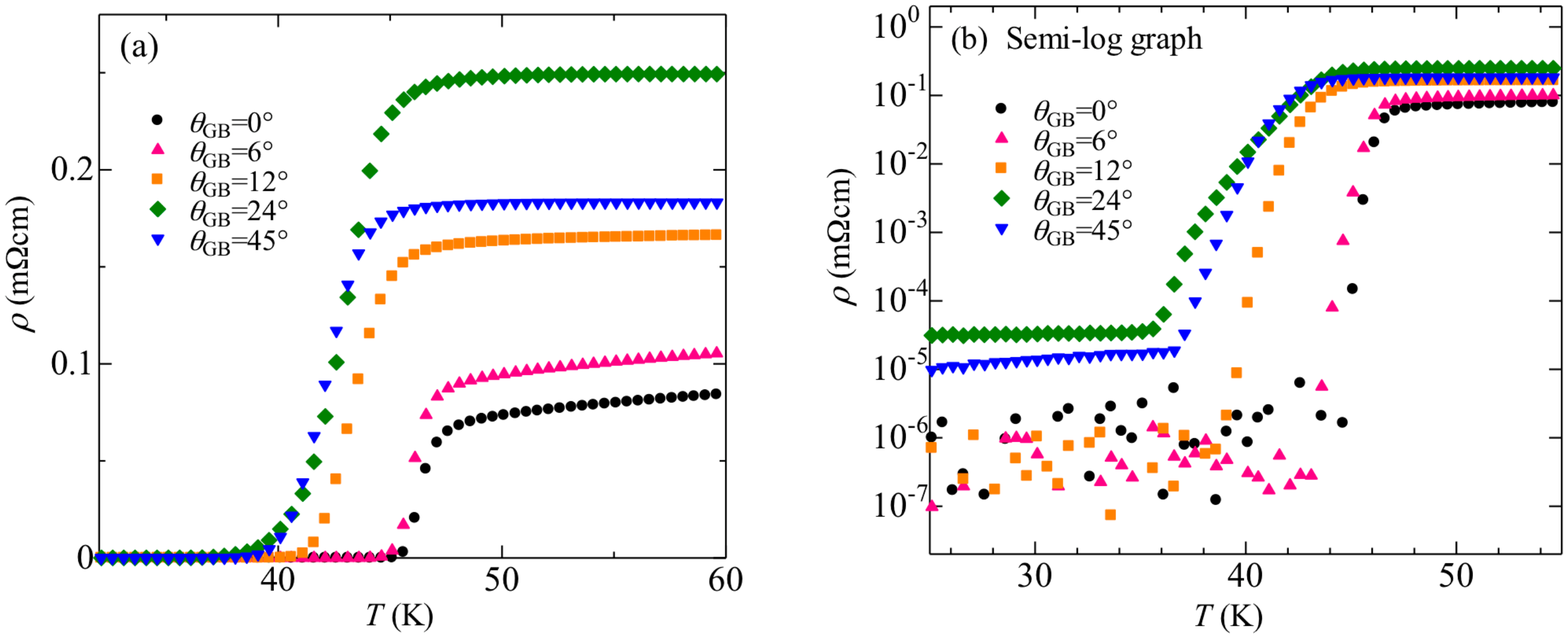}
		\caption{(a) $\rho$--$T$ curves of the inter-grain bridges with various misorientation angles and the intra-grain bridge. (b) Semi-logarithmic plots of the same $\rho$--$T$ curves.}
\label{fig:figure3}
\end{figure*}

Figure\,\ref{fig:figure2} shows the result of structural characterization by XRD of the NdFeAs(O,F) thin film on [001]-tilt MgO bicrystal with $\theta_{\rm GB}=45^\circ$. The NdFeAs(O,F) film has grown $c$--axis oriented as revealed by the $\theta$--2$\theta$ scan, Figure\,\ref{fig:figure2}(a). Impurity phase was observed as the by-product of chemical reaction between NdOF and NdFeAsO. The $\phi$--scan of the 102 NdFeAs(O,F) peak showed an eightfold pattern with a 45$^\circ$ separation as shown in Figure\,\ref{fig:figure2}(b). Hence, the NdFeAs(O,F) film has grown epitaxially on the [001]-tilt MgO bicrystal. We confirmed that all films on the bicrystal substrates with various $\theta_{\rm GB}$ were biaxially textured.

Figure\,\ref{fig:figure3} compares the resistivity curves for the inter-grain bridges with various $\theta_{\rm GB}$. Here $\theta_{\rm GB}=0^\circ$ is represented by the intra-grain bridge that was fabricated from the NdFeAs(O,F) film on the $\theta_{\rm GB}=6^\circ$ bicrystal substrate (see Figure\,\ref{fig:figure1}). The intra-grain bridge had an onset $T_{\rm c}$ ($T_{\rm c}^{\rm onset}$) of 47\,K and zero $T_{\rm c}$ ($T_{\rm c}^0$) of 46\,K. On the other hand, for inter-grain bridges, both $T_{\rm c}^{\rm onset}$ and $T_{\rm c}^0$ decreased with increasing the misorientation angle. Additionally, a finite resistivity even below $T_{\rm c}^{\rm onset}$ was observed for the inter-grain bridges with $\theta_{\rm GB}\geq24^\circ$ the as shown in Figure\,\ref{fig:figure3}(b), whereas the resistivity of the other bridges dropped well below the instrumental limitation.

\begin{figure*}
	\centering
		\includegraphics[width=12cm]{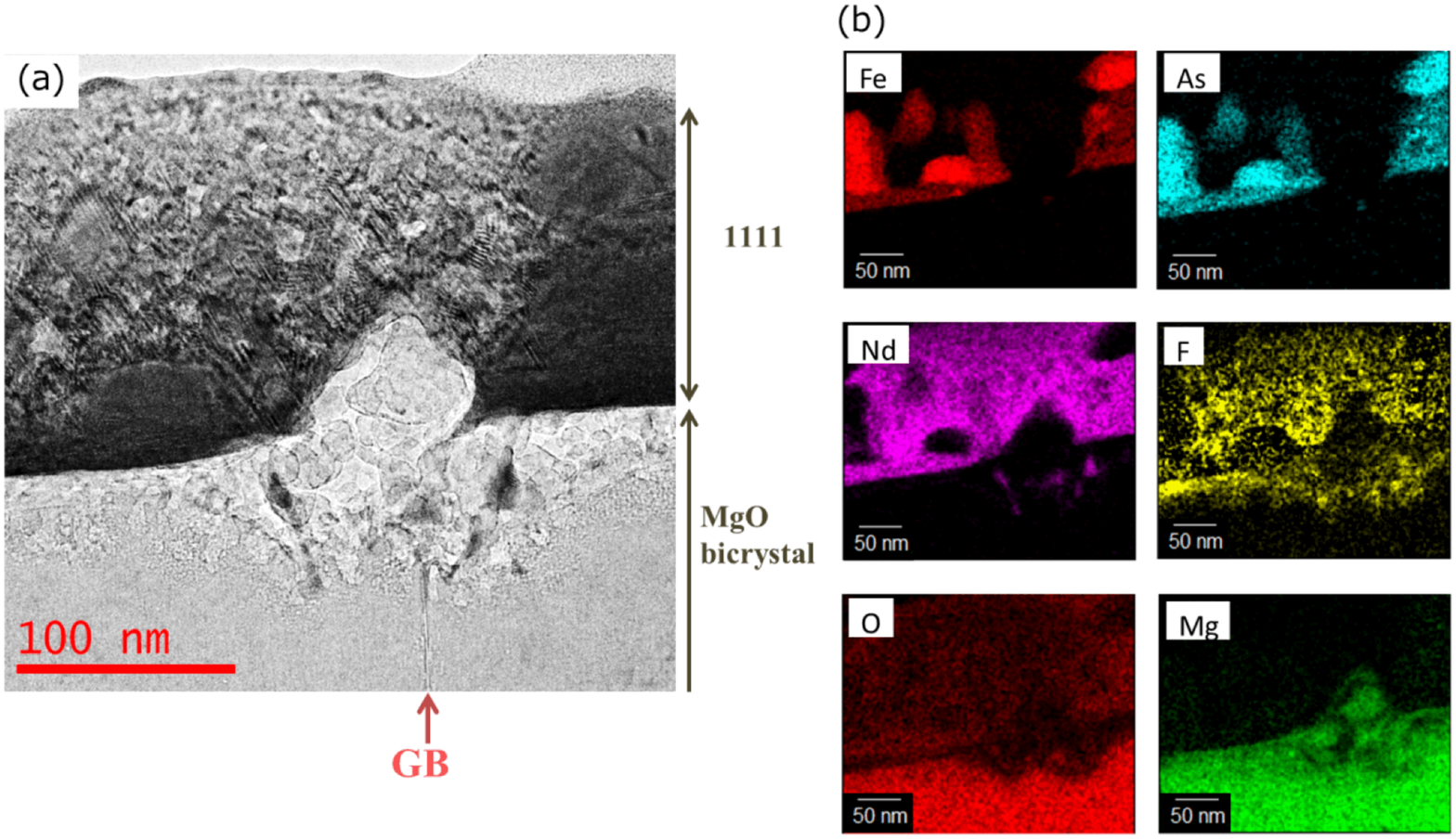}
		\caption{(a) Cross-sectional view of the NdFeAs(O,F) thin film with $\theta_{\rm GB}=24^\circ$ in the vicinity of the grain boundary region. (b) Elemental mappings acquired from the same area.}
\label{fig:figure4}
\end{figure*}

To clarify the reason why the bridges with higher misorientation angles had a finite resistivity, we performed microstructural characterization by TEM. Shown in Figure\,\ref{fig:figure4}(a) is the cross-sectional view of the NdFeAs(O,F) thin film with $\theta_{\rm GB}\geq24^\circ$ in the vicinity of GB region. Clearly, both the NdFeAs(O,F) layer and the MgO substrate around GB are damaged. Elemental mappings acquired from the same area showed the absence of Fe and As, and the segregation of Mg (Figure\,\ref{fig:figure4}(b)). Other than the GB region, we did not see such damage to the NdFeAs(O,F) film and the substrate. These results show that F preferentially diffused through GB and thereby eroded NdFeAs(O,F) and MgO. It was reported for a YBCO/Ca-doped YBCO bilayer film that Ca diffused from the over-layer into the grain boundary region of the bottom YBCO layer, causing hole overdoping and consequently increased the inter-grain $J_{\rm c}$.\cite{Hammerl} Similarly, fluorine had probably diffused more easily along the grain boundary in our films, but unlike YBCO, the excess fluorine gave a negative impact on the structural and transport properties because of its strong reactivity.

\begin{figure*}
	\centering
		\includegraphics[width=12cm]{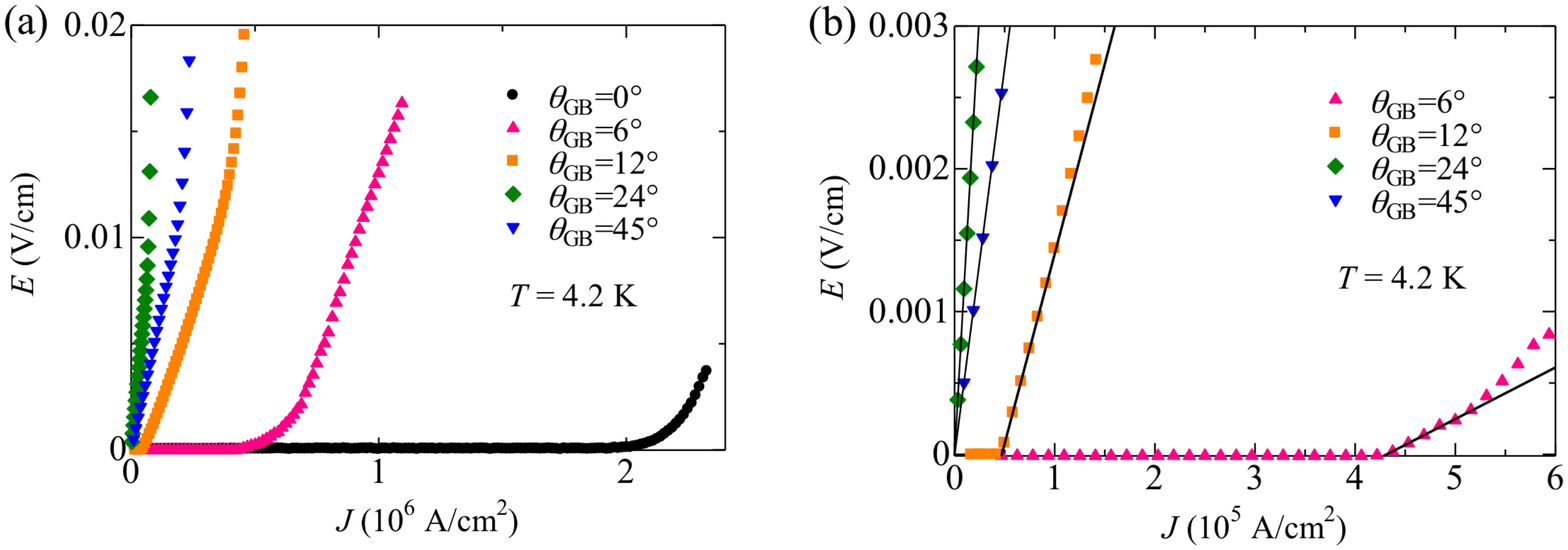}
		\caption{(a) $E$--$J$ characteristics of the inter-grain bridges for various misorientation angles and the intra-grain bridge. (b) Enlarged view of the $E$--$J$ curves at low $E$.}
\label{fig:figure5}
\end{figure*}

Shown in Figure\,\ref{fig:figure5}(a) is the $E$--$J$ characteristics of the inter-grain bridges for various $\theta_{\rm GB}$ measured at 4.2\,K. For comparison, the data of the intra-grain bridge is also plotted. As can be seen, the intra-grain curve showed a typical power-law behavior (i.e., $E$$\sim$$J^n$), suggesting that $J_{\rm c}$ is limited by the intra-grain itself. On the other hand, the inter-grain bridges for $\theta_{\rm GB}=6^\circ$ and 12$^\circ$ showed a linear behavior at low $E$ due to flux flow along the GB, indicative of $J_{\rm c}$ limitation by GB (Figure\,\ref{fig:figure5}(b)).\cite{Diaz, Verebelyi} With further increasing $\theta_{\rm GB}$, a finite resistivity was observed as shown in Figure\,\ref{fig:figure3}(b), and accordingly, the inter-grain $J_{\rm c}$ could not be evaluated.

\begin{figure}
	\centering
		\includegraphics[width=\columnwidth]{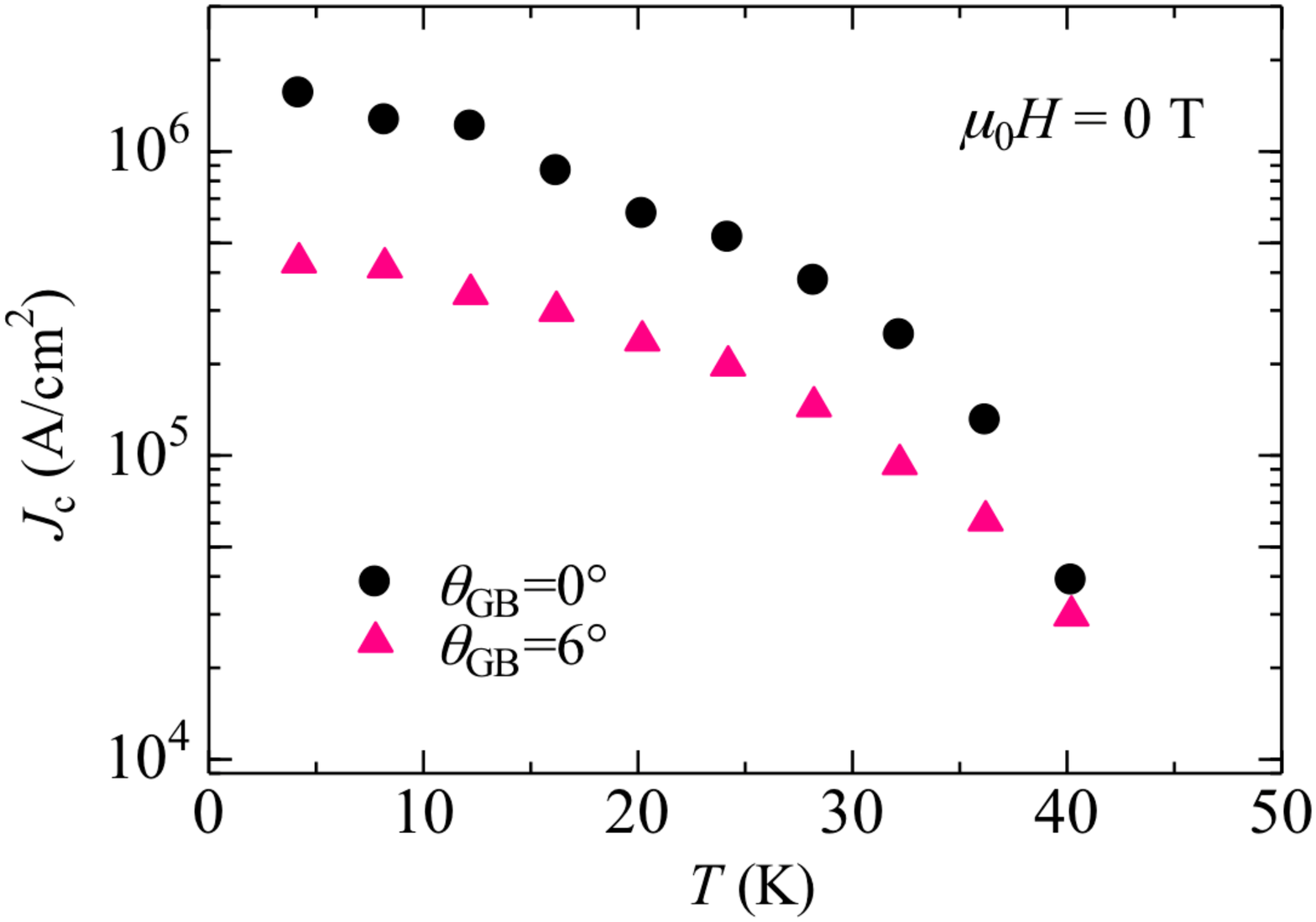}
		\caption{Self-field intra- and inter-grain ($\theta_{\rm GB}=6^\circ$) $J_{\rm c}$ as a function of temperature.}
\label{fig:figure6}
\end{figure}

Figure\,\ref{fig:figure6} shows the inter-grain $J_{\rm c}$ for $\theta_{\rm GB}=6^\circ$ measured at various temperatures without applying a magnetic field. The intra-grain $J_{\rm c}$ of the same film is also plotted for comparison. At low temperatures, the self-field $J_{\rm c}$ of the intra-grain exceeded 1\,MA/cm$^2$. However, $J_{\rm c}^{\rm inter}$ reduced by almost 30\% compared to $J_{\rm c}^{\rm intra}$. Hence, the critical angle is less than 9$^\circ$, which is different from Co-doped BaFe$_2$As$_2$ and Fe(Se,Te).

\begin{figure}
	\centering
		\includegraphics[width=\columnwidth]{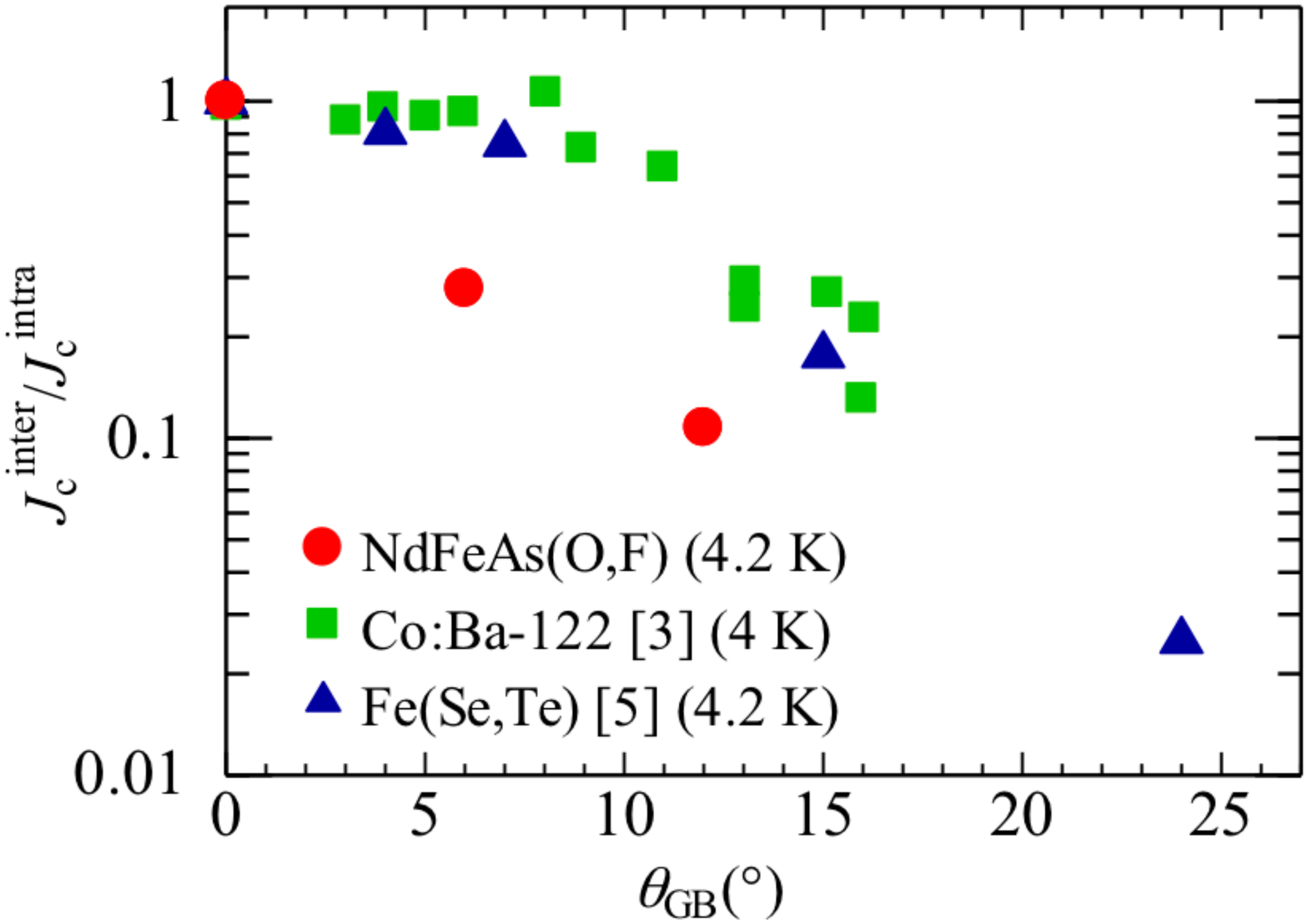}
		\caption{The ratio of inter- and intra-grain  $J_{\rm c}$ ($J_{\rm c}^{\rm inter}/J_{\rm c}^{\rm intra}$) as a function of $\theta_{\rm GB}$. For comparison, the data of Co-doped BaFe$_2$As$_2$\,\cite{Katase} and Fe(Se,Te)\,\cite{Si} are also shown.}
\label{fig:figure7}
\end{figure}

Figure\,\ref{fig:figure7} summarizes the ratio of inter- and intra-grain $J_{\rm c}$ as a function of misorientation angle. For comparison, the data for Co-doped BaFe$_2$As$_2$\,\cite{Katase} and Fe(Se,Te) \,\cite{Si} are also plotted. As can be seen, the ratio is nearly unity until $\theta_{\rm GB}$$\sim$9$^\circ$ for both Co-doped BaFe$_2$As$_2$ and Fe(Se,Te), indicative of strong-coupling between the grains. On the other hand, our NdFeAs(O,F) film showed a weak link behavior even at $\theta_{\rm GB}=6^\circ$. However, the present result may not reflect the intrinsic GB properties of NdFeAs(O,F). As stated above, the inter-grain bridges for $\theta_{\rm GB}\geq24^\circ$ had a finite resistivity due to the erosion of GB region by F. Although we observed a non-zero $J_{\rm c}$ for $\theta_{\rm GB}=6^\circ$ and 12$^\circ$, it is well likely that fluorine had diffused in excess along the GB region in these films as well and caused some damages. Hence the present result of inter-grain $J_{\rm c}$ is probably an underestimation.

To characterize the intrinsic nature of grain boundary for NdFeAs(O,F), the excess F-diffusion to GB should be prevented. In our earlier studies, the growth temperature was 670$^\circ$C for both NdFeAsO and NdOF. \cite{Kawaguchi-3,Uemura} Later we found that increasing the growth temperature improved the crystallinity of the 1111 phase,\cite{Chihara} and we have grown both the 1111 and NdOF phases at 800$^\circ$C in the present study. However, the growth temperature of NdOF over-layer would not influence the crystallinity of the 1111 layer. Hence, suppressing the excess diffusion of fluorine without compromising the crystallinity of the 1111 phase might be achieved by lowering the growth temperature of NdOF whilst keeping that of NdFeAsO at 800$^\circ$C, which would be the direction of a future study. That method would also be useful to improve the performance of 1111-based coated conductors. Previously, we reported the field ($H$) dependence of $J_{\rm c}$ for NdFeAs(O,F) coated conductors.\cite{Iida} Although $T_{\rm c}$ was around 43\,K, almost twice as high as that of Co-doped BaFe$_2$As$_2$, the $J_{\rm c}$--$H$ performance was inferior to that of Co-doped BaFe$_2$As$_2$. However, this NdFeAs(O,F) coated conductor was fabricated at 800$^\circ$C. Based on the results of the present study, we think that the grain boundaries of the coated conductor had suffered damages from excess fluorine, and the suppression of excess F-diffusion would improve the $J_{\rm c}$--$H$ properties.

\section{Summary}
NdFeAs(O,F) thin films were grown on 6$^\circ$, 12$^\circ$, 24$^\circ$ and 45$^\circ$ [001]--tilt MgO bicrystal substrates and their electrical transport properties were investigated. Structural characterization by X-ray diffraction revealed that all films have grown epitaxially. Finite resistivity was observed for the inter-grain bridges with higher misorientation angles ($\theta_{\rm GB}\geq24^\circ$). Additionally, by comparing with Co-doped BaFe$_2$As$_2$ and Fe(Se,Te) grain boundary junctions, the inter-grain $J_{\rm c}$ showed a larger decay with $\theta_{\rm GB}$. As a possible reason of this result, fluorine may have diffused preferentially to the grain boundary region and degraded the structural and transport properties. The suppression of excess F-diffusion to GB would therefore be the key for studying the intrinsic grain boundary properties.

\begin{acknowledgments}
This work was partially supported by the JSPS Grant-in-Aid for Scientific Research (B) Grant Number 16H04646. We acknowledge Kimitaka Higuchi of High Voltage Electron Microscope Laboratory, Institute of Materials and Systems for Sustainability, Advanced Characterization Nanotechnology Platform, for TEM observation.
\end{acknowledgments}

\end{document}